\documentclass[fleqn,12pt,twoside]{article}
\usepackage[headings]{espcrc1}

\readRCS
$Id: espcrc1.tex,v 1.2 2004/02/24 11:22:11 spepping Exp $
\usepackage{graphicx}
\usepackage[figuresright]{rotating}

\newcommand{\AmS}{{\protect\the\textfont2
  A\kern-.1667em\lower.5ex\hbox{M}\kern-.125emS}}
\title{QCD matter within a quasi-particle model and the critical end point}

\author{B. K\"ampfer\address[FZR]{Forschungszentrum Rossendorf, 
                     Institut\ f\"ur Kern- und Hadronenphysik,\\ PF 510119, 
                     01314 Dresden, Germany}\address[TUD]{Technische Universit\"at Dresden, 
                     Institut f\"ur Theoretische Physik,\\
                     01062 Dresden, Germany},
        M. Bluhm\addressmark[FZR]\thanks{Supported by BMBF and  GSI-FE.},
        R. Schulze\addressmark[TUD],
        D. Seipt\addressmark[TUD]
        and
        U. Heinz\address[Ohio]{Ohio State University, 
        Department of Physics, Columbus, OH 43210, USA}}

\runtitle{QCD matter within a quasi-particle model}
\runauthor{B. K\"ampfer}

\begin{document}

% typeset front matter
\maketitle

\begin{abstract}
We compare our quasi-particle model with recent lattice QCD results for the
equation of state at finite temperature and baryo-chemical potential.
The inclusion of the QCD critical end point
into models is discussed.
We propose a family of equations of state to be employed in hydrodynamical
calculations of particle spectra at RHIC energies and compare
with the differential azimuthal anisotropy of strange and charm hadrons.   
\end{abstract}

\section{QUASI-PARTICLE MODEL}

The equation of state (EoS) of strongly interacting matter is a central issue
to understand the hydrodynamical evolution in high-energy heavy-ion collisions.
{\it Ab initio} evaluations based on QCD can deliver the EoS, e.g., 
in the form \cite{Allton03}
\begin{equation}
p = %-B + 
T^4 \left\{ c_0(T) + 
c_2(T) \left( \frac{\mu}{T} \right)^2 + 
c_4(T) \left( \frac{\mu}{T} \right)^4 + 
c_6(T) \left( \frac{\mu}{T} \right)^6 + \cdots \right\}.
\end{equation}
%There is no hint to finite values of the first term, therefore $B = 0$.
To arrive at a flexible parametrization of the EoS we have developed
a quasi-particle model \cite{Peshier,Bluhm05}
using dynamically generated self-energies and a non-perturbative 
effective coupling as essential input. A comparison of the model with
recent lattice data is exhibited in Fig.~1. 
The peak of $c_4$ and the dipole shape of $c_6$ emerge from a change
of the effective coupling at the pseudo-critical temperature $T_c(\mu_B = 0)$
dictated by the data.
The calculation of the coefficients $c_{2,4,6}$ requires the knowledge of the
pressure $p(T,\mu_B)$ at finite values of baryo-chemical potential
$\mu_B$ which is delivered by
Peshier's equation \cite{Peshier}. The baryon number susceptibility
follows directly from $p(T, \mu_B)$, see Fig.~1. % right bottom panel.
The peak developing for $\mu_B > 300$ MeV can be considered as a signal 
of some critical behavior.
\begin{figure}[htb]
\includegraphics[angle=-90,width=16pc]{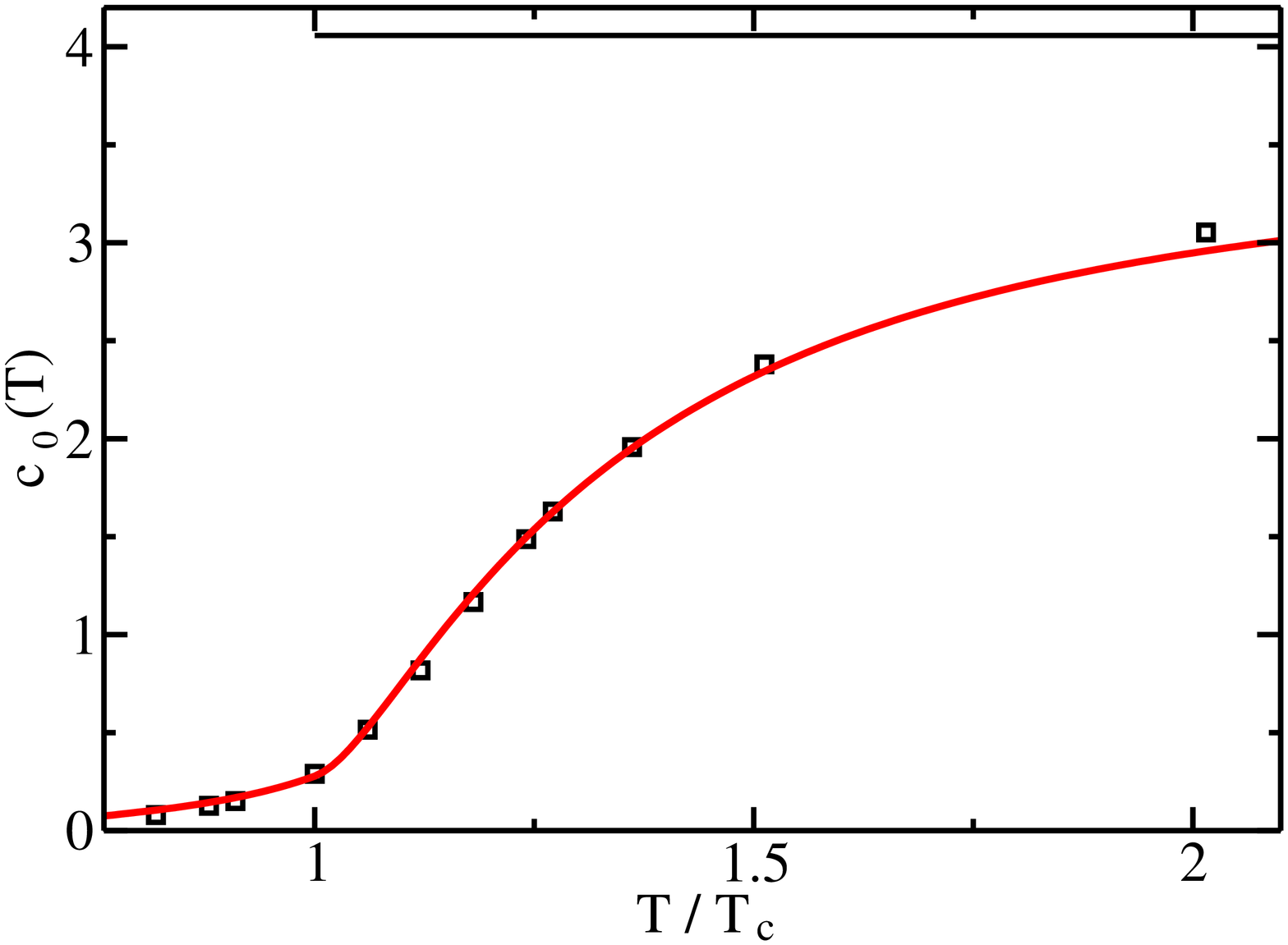} \hfill
\includegraphics[angle=-90,width=16pc]{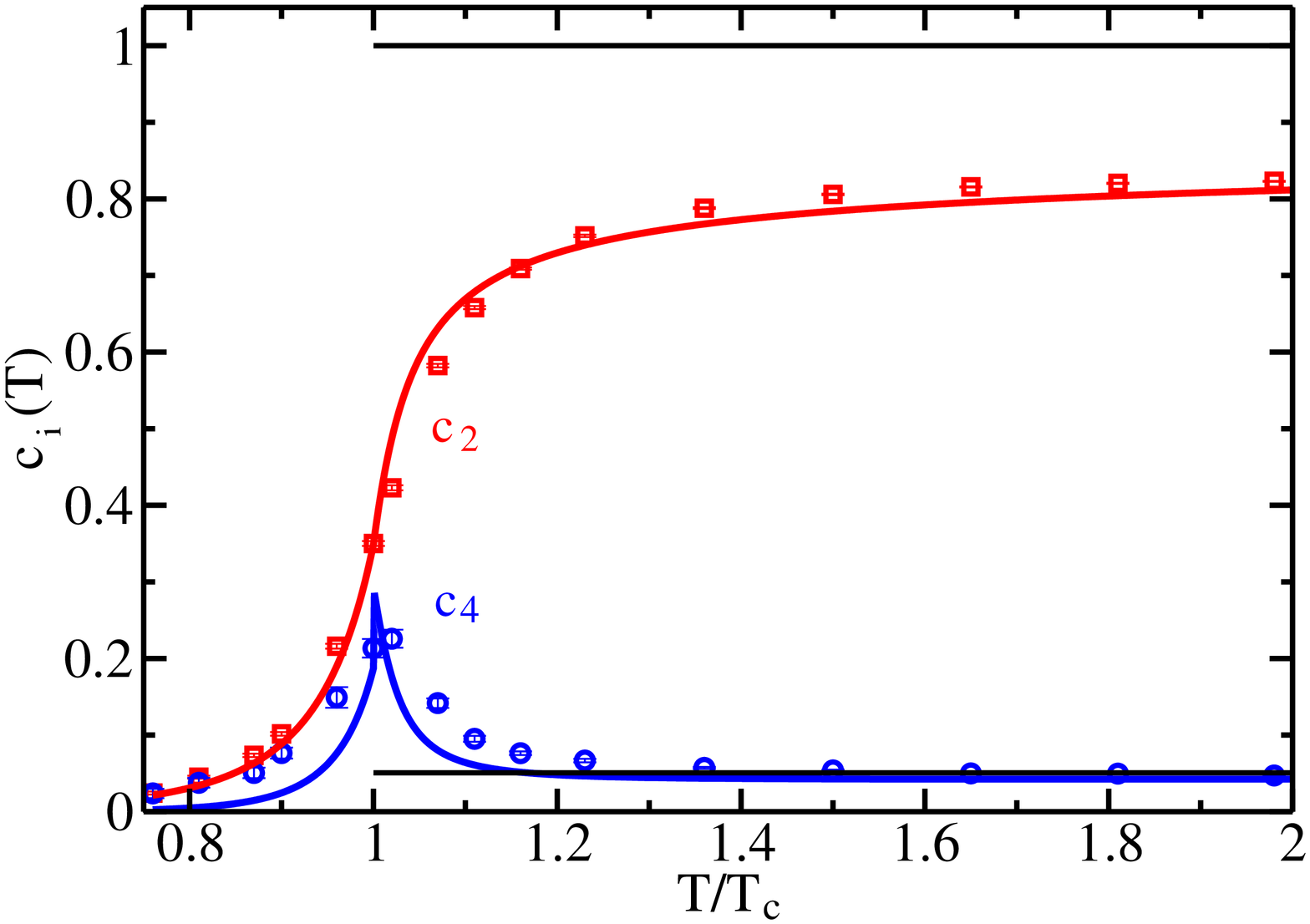}\\[4mm]
\includegraphics[angle=-90,width=16pc]{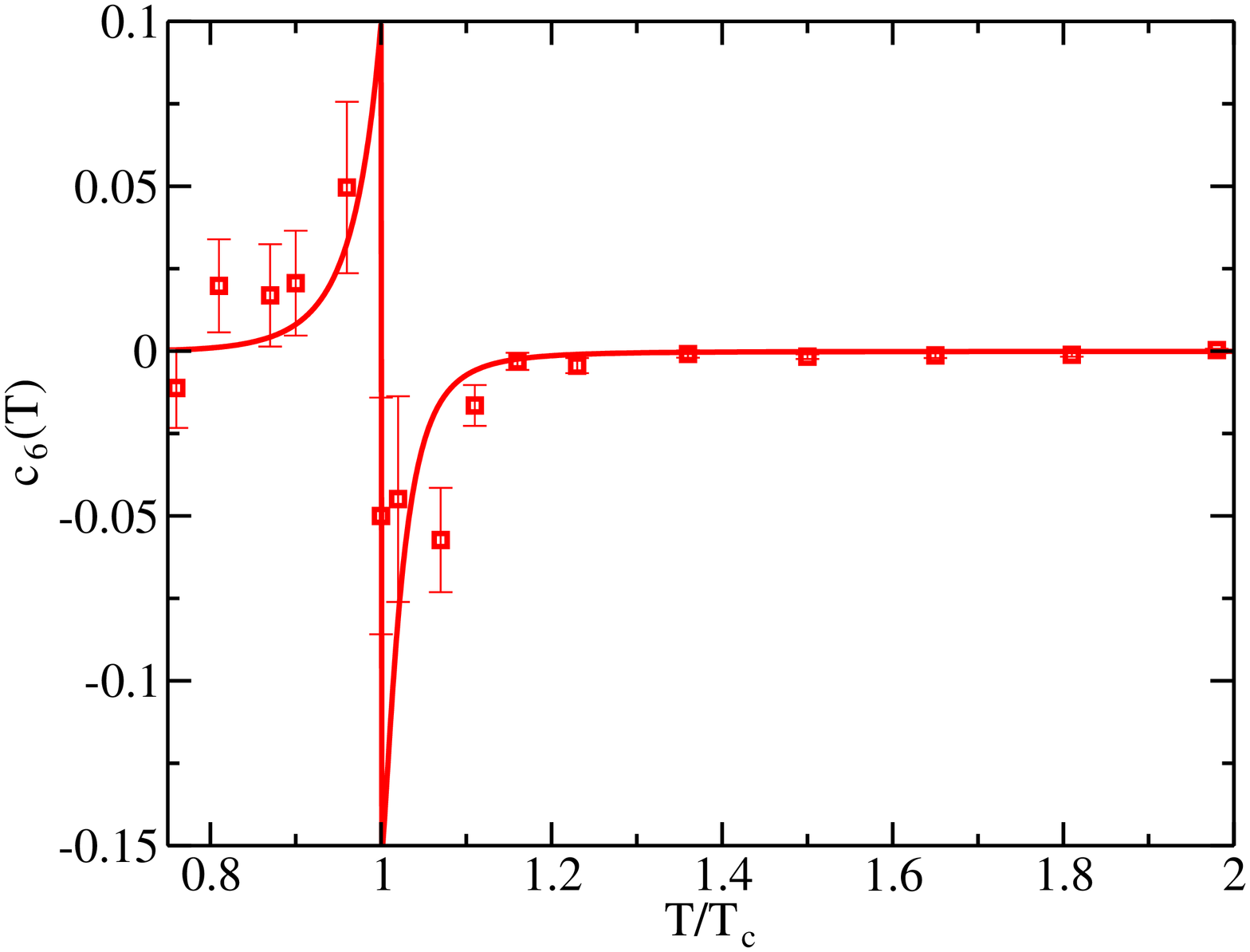} \hspace*{1.7cm}
\includegraphics[angle=-90,width=17pc]{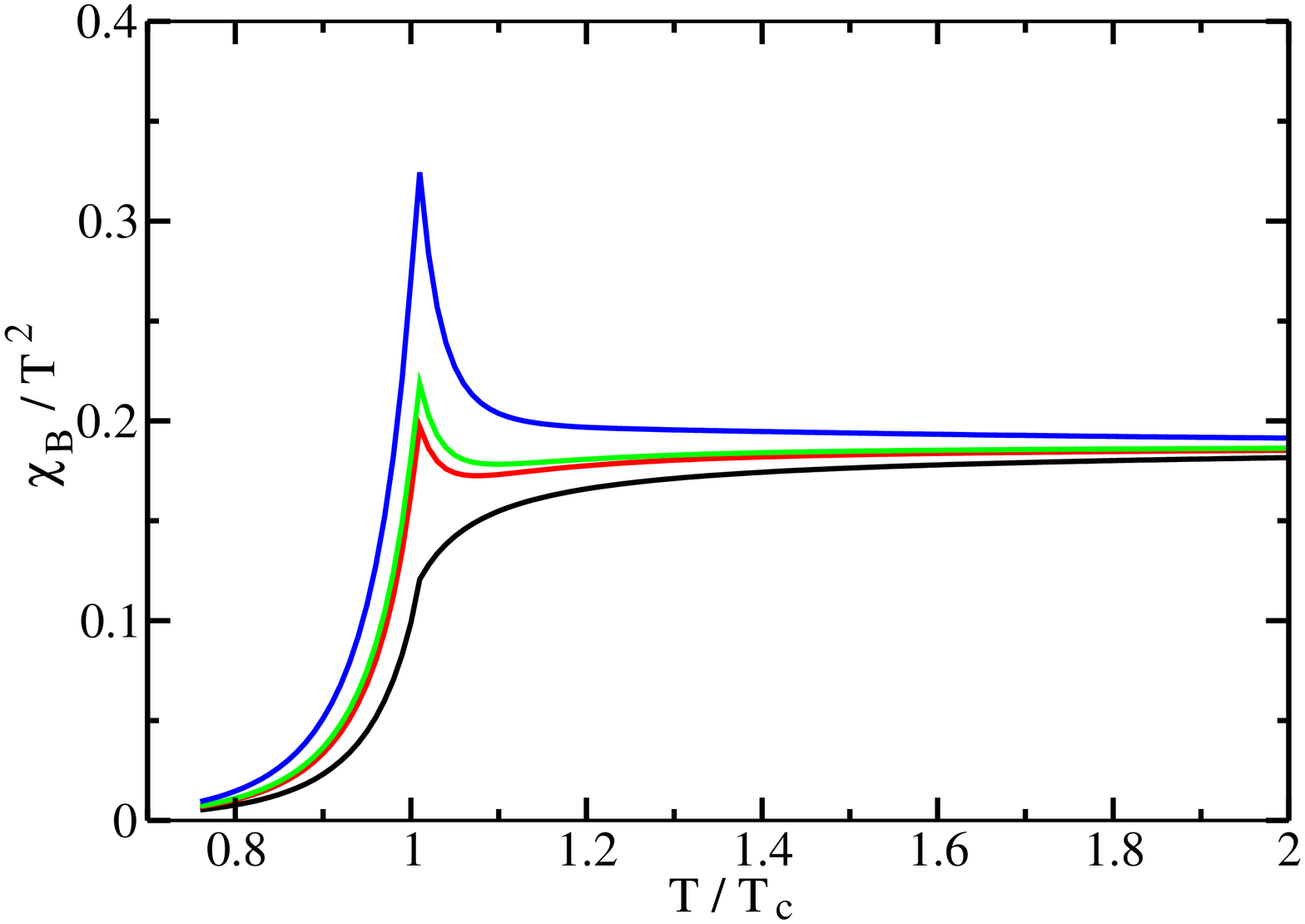}
\vskip -3mm
\caption{The Taylor expansion coefficients $c_{0, 2, 4}$ (top row,
horizontal lines depict the corresponding terms in the bag model)
and $c_6$ (left panel in bottom row) 
of the equation of state as a function of the scaled temperature.
QCD lattice data from \cite{Allton03,Laermann}.
Right panel in bottom row: baryon number susceptibility 
($\mu_B =$ 450, 330, 300, 150 MeV from top to bottom).
For details see \cite{Bluhm05}.}
\label{fig:ci}
\end{figure}

\section{CRITICAL END POINT}

QCD for finite quark masses displays a critical end point (CEP) \cite{CEP} 
where a phase transition of first order sets in at finite values of $\mu_B$. 
The CEP belongs to the
universality class of the 3D Ising model characterized by a set of
critical exponents. A convenient parametric
representation of the singular part (formulated in terms of the entropy
density $s$) of the EoS can be found in \cite{Zinn_Justin}.
The complete EoS is accordingly
\begin{equation}
s = s_{reg} + s_{sing}.
\end{equation}
While the location of the CEP is determined by lattice QCD calculations
\cite{Fodor} the size of the critical region is barely known.
A simple toy model demonstrating the effect of $s_{sing}$ for given
$s_{reg}$ is $s_{reg} = 4 \bar c_0 T^3 + 2 \bar c_2 \mu_B^2 T$ and
$s_{sing} = s_{reg} A \tanh {\cal S}_c$ with the same ${\cal S}_c$ as in
\cite{Asakawa} and $\bar c_0 = (32+21 N_f)\pi^2/180$ and $\bar c_2 = N_f/18$
for $N_f = 2$.
We assume a critical line given by 
$T(\mu_B) = T_c (1 - c (\mu_B/T_c)^2)$ with $c = 0.0077$
from \cite{Allton02}
and locate the CEP at $\mu_B = 330$ MeV according to \cite{Fodor}. 
This information is needed to map the coordinates of $s_{sing}$ into the $T - \mu_B$ plane.
A typical pattern of isentropic trajectories is exhibited
in Fig.~2. The pattern (left panel) seems to be generic 
(see also \cite{Asakawa} for another model with CEP):
trajectories which are left [right] to the CEP for 
small values of the strength parameter $A$ are attracted [repelled],
but at variance to lattice QCD results (right panel). 

The proper implementation of CEP phenomena in our quasi-particle model
is under consideration.

\begin{figure}[ht]
\begin{minipage}[t]{9.3cm}
\includegraphics[angle=0,width=15pc]{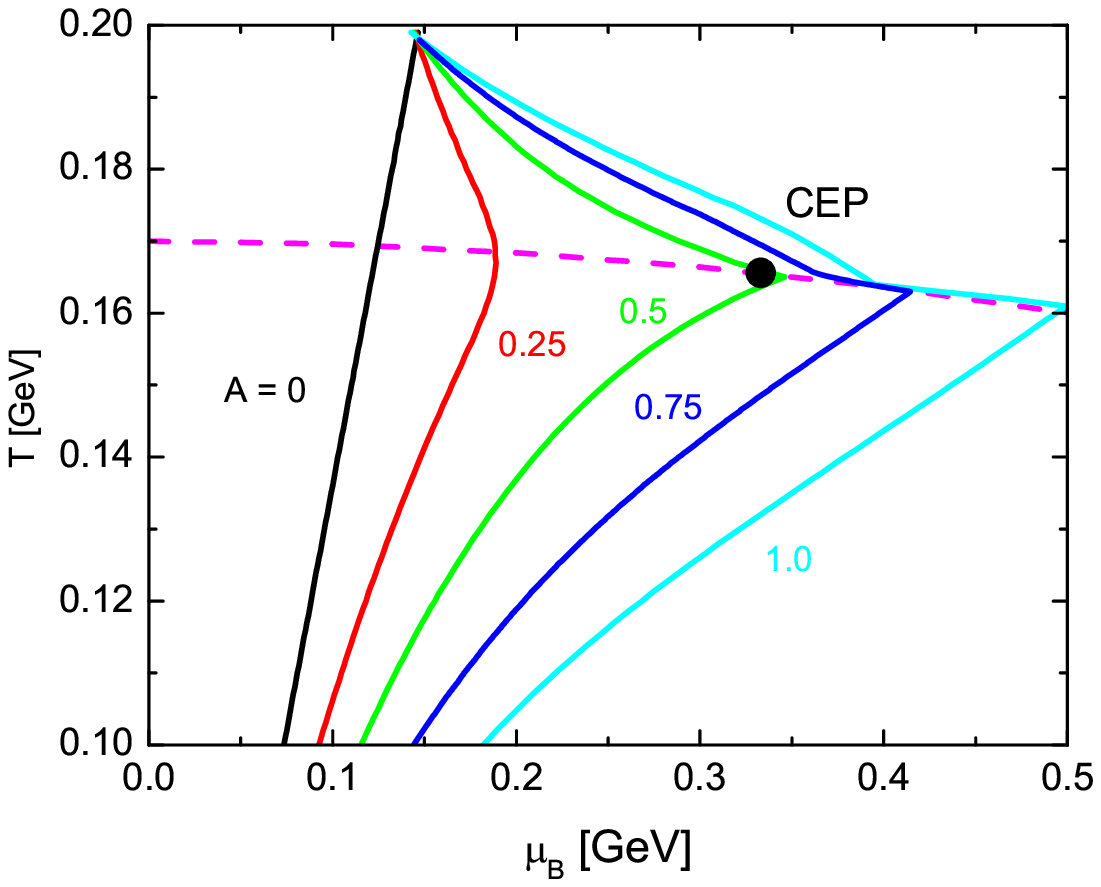}
\end{minipage}
\hfill
\begin{minipage}[t]{8cm}
\vskip -3.7mm
\includegraphics[angle=-90,width=15pc]{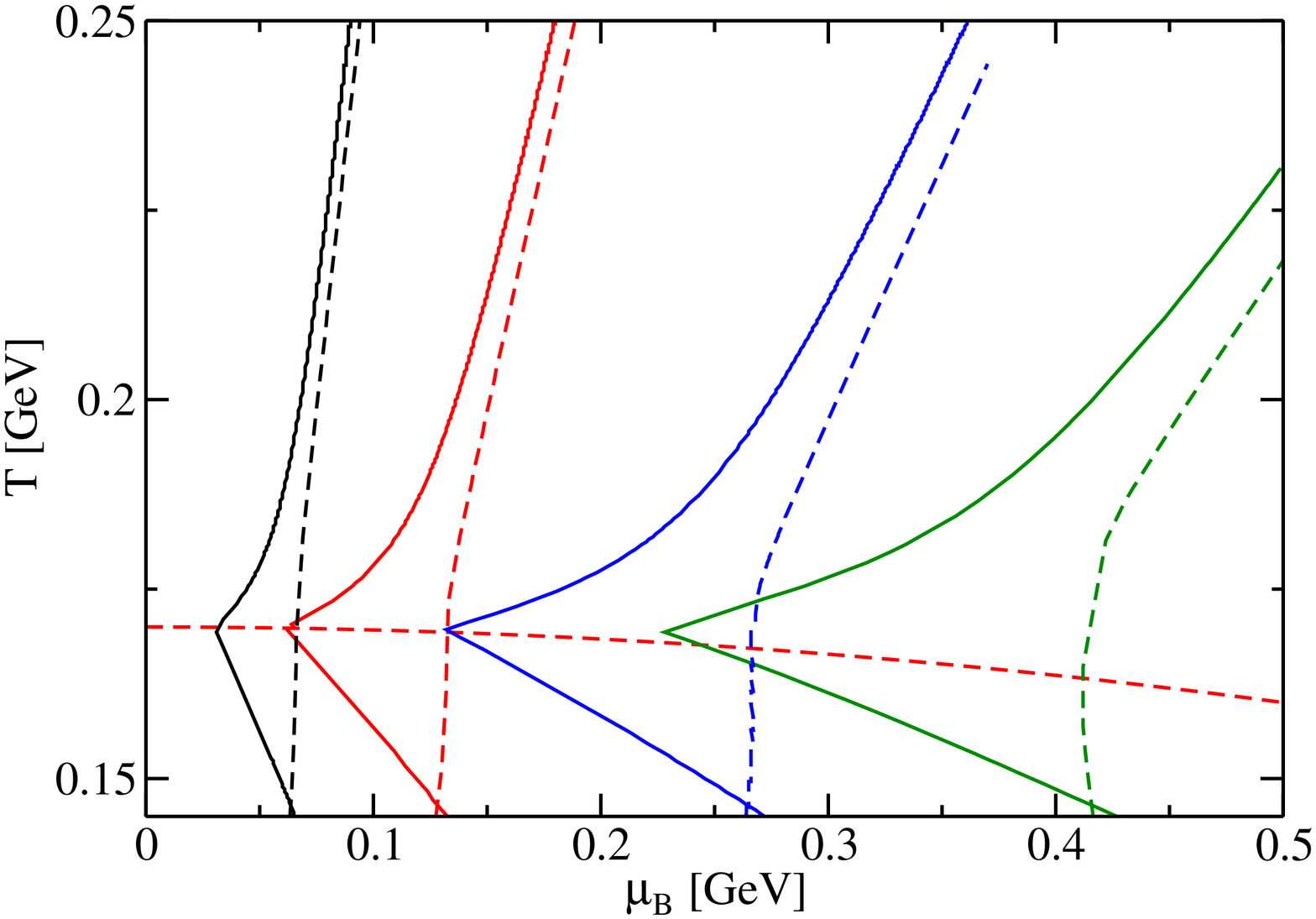}%{c2trans10.eps}
\end{minipage}
\vskip -6mm
\caption{Isentropic trajectories and the (pseudo-)critical line 
in the $T - \mu_B$ plane.
Left panel: a toy model including the CEP 
(entropy per baryon = 100, various strength parameters $A$ as indicated).
Right panel: quasi-particle model 
adjusted to lattice QCD data \cite{Allton03,Laermann} via
$c_0$ (solid curves) or $c_2$ (dashed curves); both fits agree
within 10\% for $p(T, \mu_B =0)$;
entropy per baryon = 200, 100, 50, 33, from left to right.}
\label{fig:isentropes}
\end{figure}

\section{A FAMILY OF EQUATIONS OF STATE, HYDRODYNAMICS AND\\
AZIMUTHAL ANISOTROPY}

It is expected that lattice QCD data at high temperature are realistic,
as improved actions have been used. At low temperature, the employed quark masses
are too large to give realistic results. These however agree with the resonance
gas model once analog assumptions are implemented \cite{Redlich}. It is therefore
reasonable to use the resonance gas model at low $T$ and employ
our quasi-particle model to extrapolate lattice QCD results both to
larger $\mu_B$ and smaller $T$ as long as 
a systematic chiral extrapolation can not be done safely 
due to lacking lattice QCD input.\\[-12mm]
\begin{figure}[hb]
\includegraphics[angle=0,width=16pc]{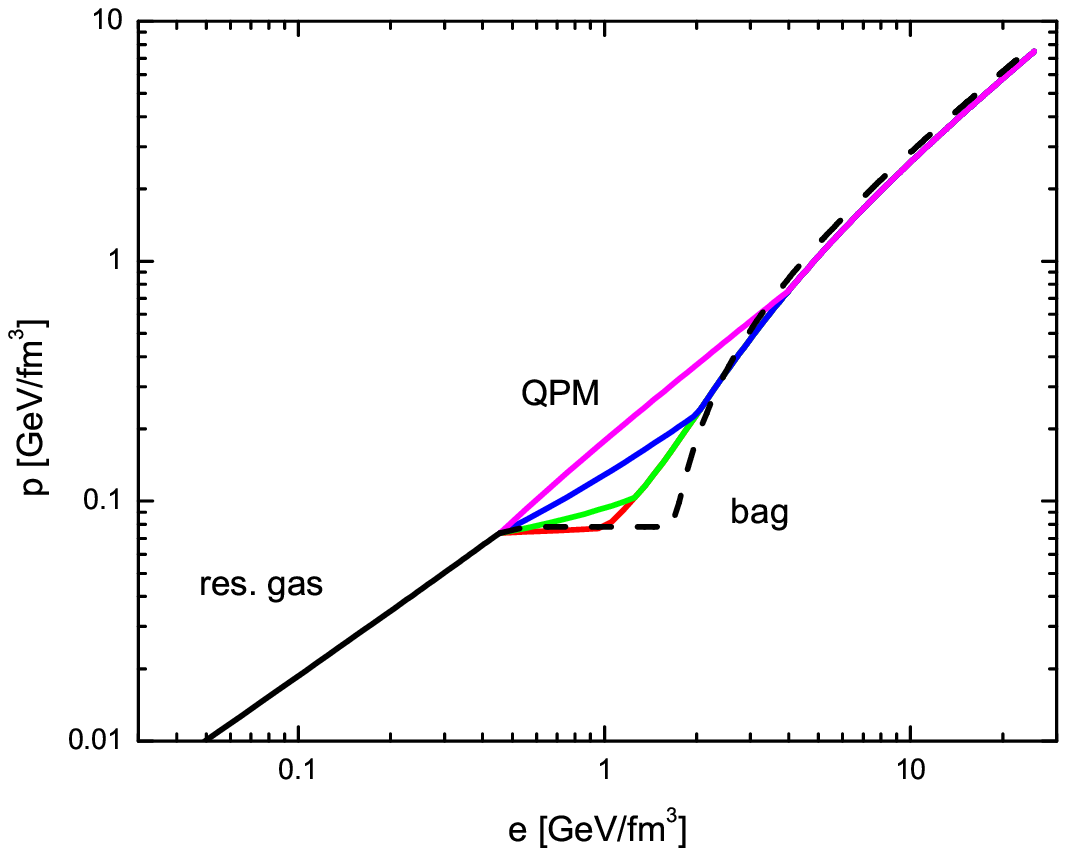} \hfill
\includegraphics[angle=0,width=16.3pc]{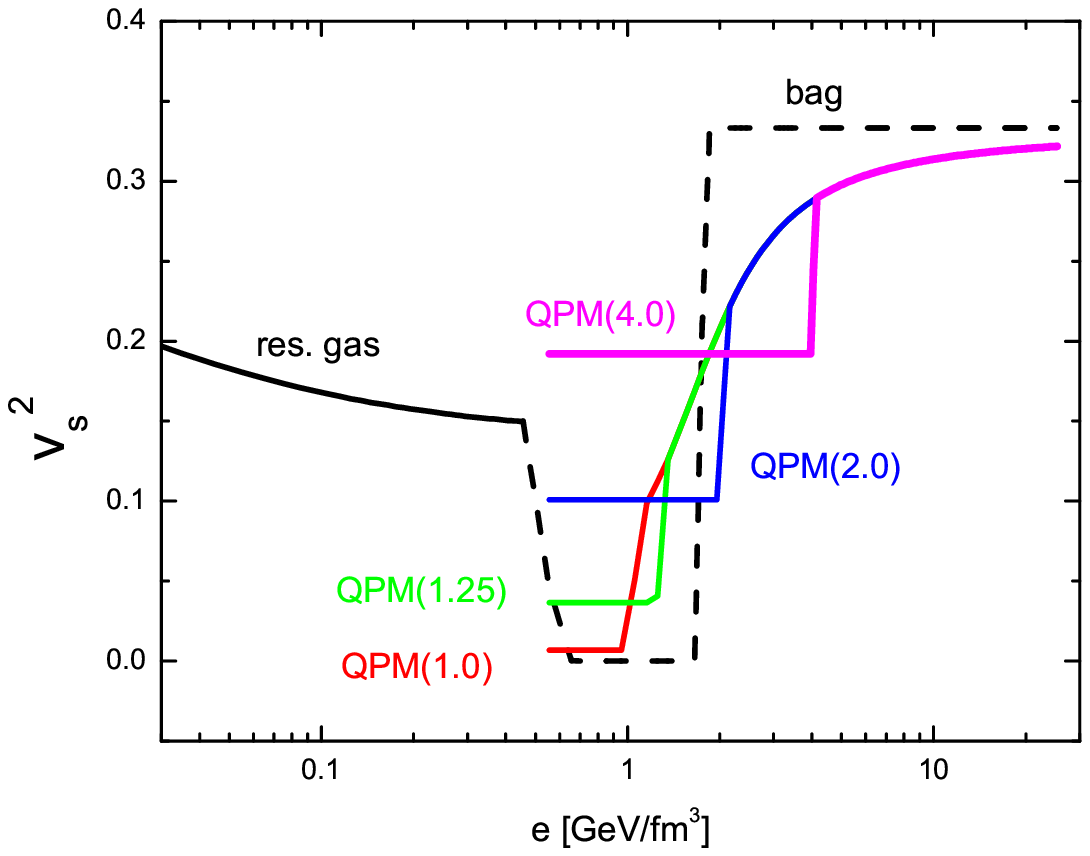}
\vskip -6mm
\caption{A family of EoS's
by combining our quasi-particle model and the resonance gas model
for baryon densities relevant for top RHIC energies:
pressure (left, QPM(4.0), QPM(2.0), QPM(1.25), QPM(1.0), from top to bottom) 
and squared sound velocity (right) as a function of energy density.
Bag model results are depicted by dashed curves.}
\label{fig:EoS}
\end{figure}

\begin{figure}[htb]
\vskip -6mm
\begin{minipage}[b]{90mm}
\includegraphics[angle=0,width=21pc]{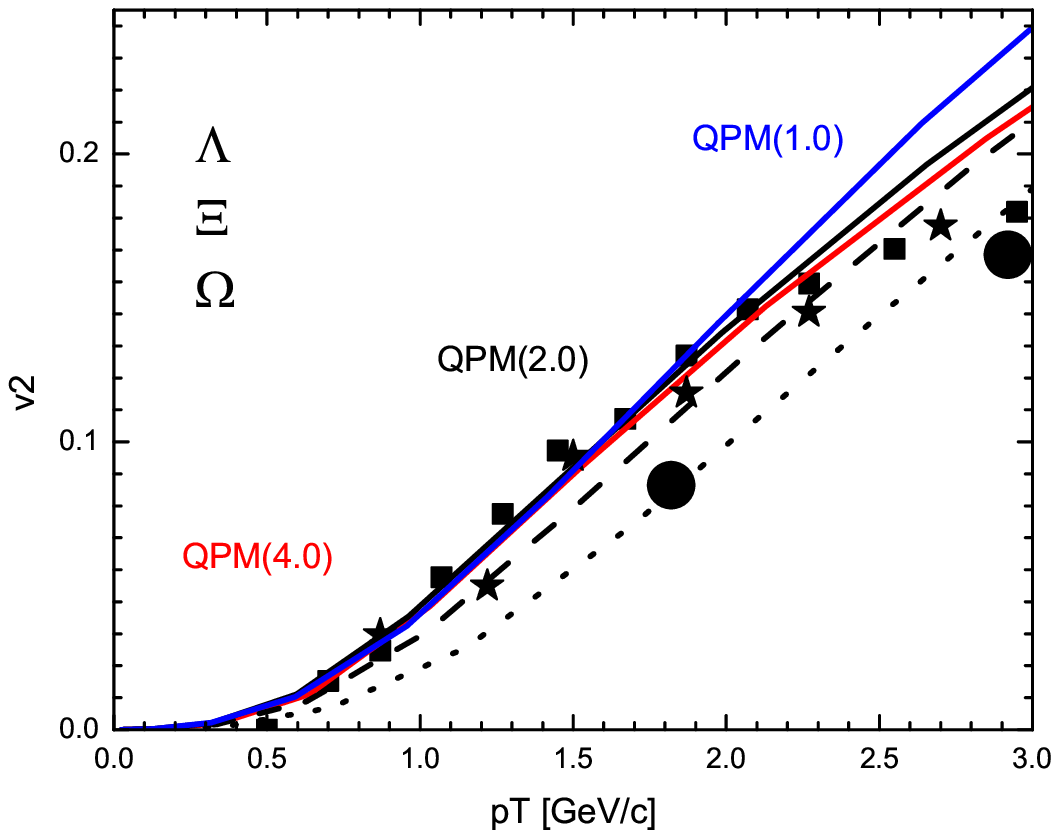}
\end{minipage}
\hfill
\begin{minipage}[b]{55mm}
\hspace*{-0.8mm}
\includegraphics[angle=0,width=11.5pc]{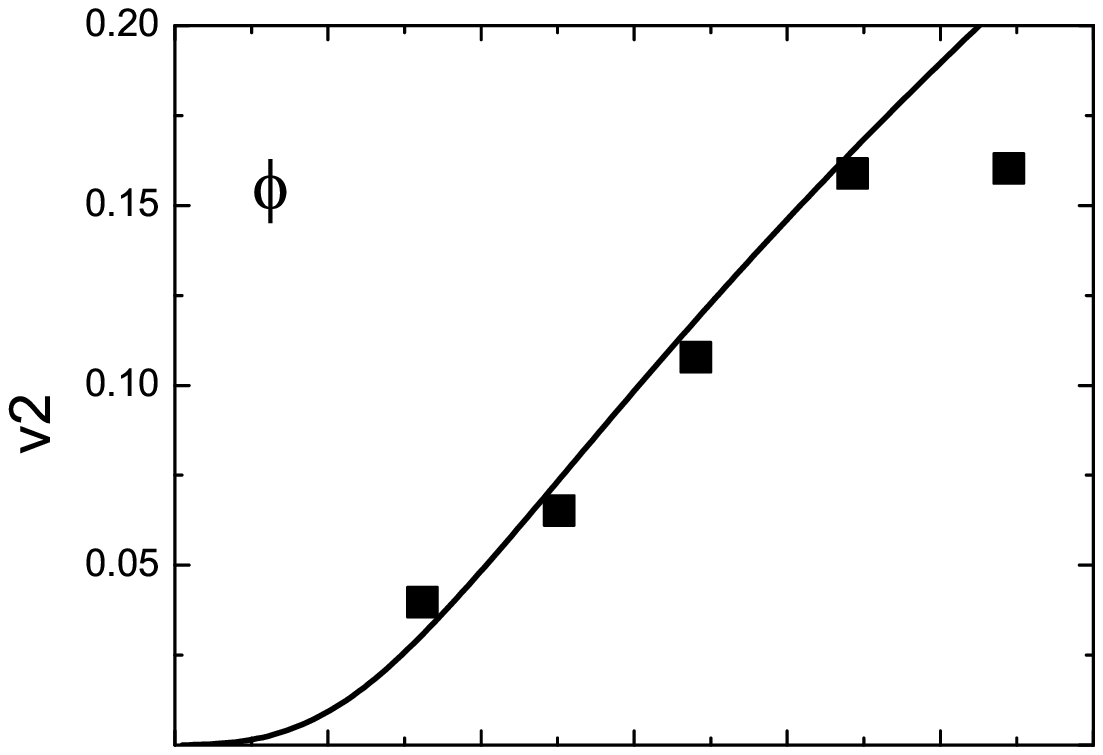}\\[-3.1mm]
\includegraphics[angle=0,width=11.5pc]{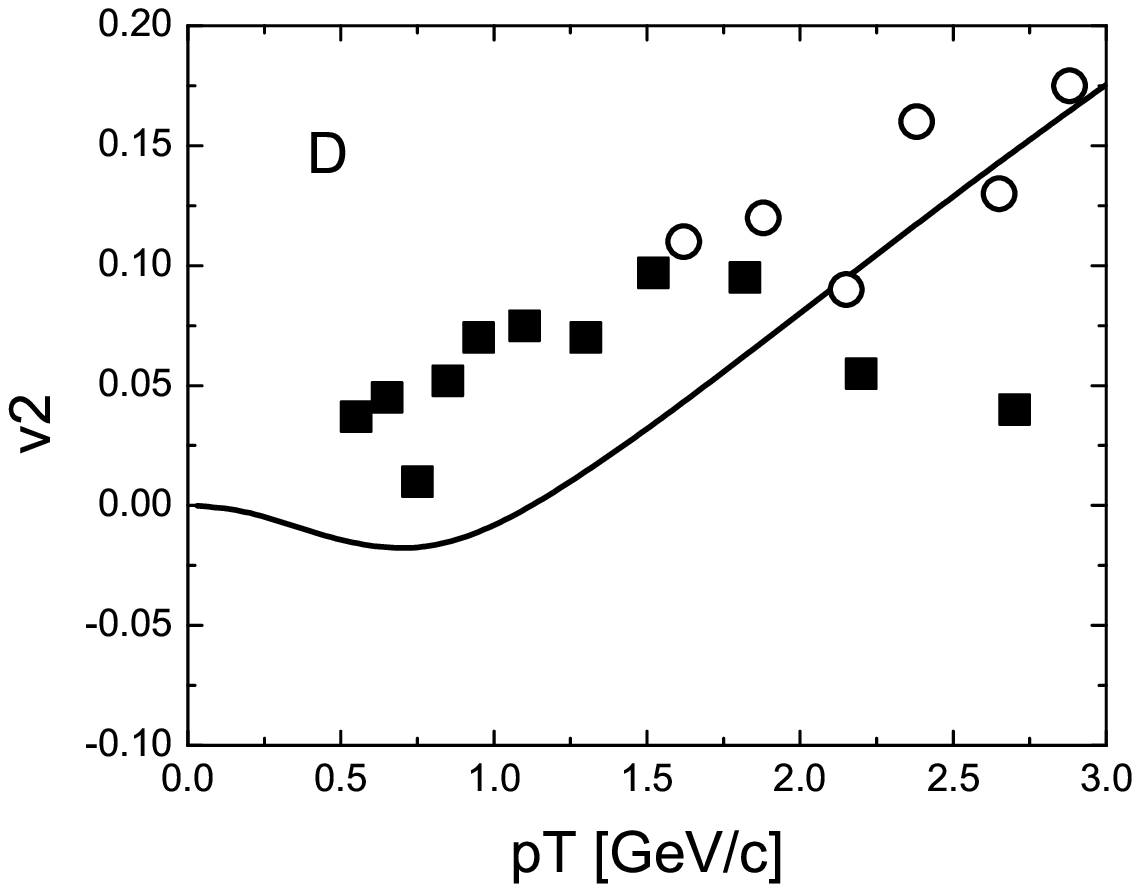}\\[-1mm]
\end{minipage}
\vskip -7.5mm
\caption{$v_2$ as a function of transverse momentum.
Left: strange baryons (solid curves QPM(1.0), QPM(2.0), QPM(4.0) [from top to bottom, 
with decoupling temperatures adjusted
to data up to 2.5 GeV/c] and squares: $\Lambda$,
dashed curve [QPM(2.0] and stars: $\Xi$, dotted curve [QPM(2.0)] and circles: $\Omega$,
data from \cite{strange_baryons}), 
right top: $\phi$ (data from \cite{phi}), 
right bottom: $D$ (data from \cite{D_meson}), both for QPM(2.0).
Impact parameter $b = 5.2$ fm.}
\label{fig:v2}
\end{figure}

We generate a family of EoS's by keeping the matching point to the resonance gas EoS
fixed and interpolate linearly to a given matching point (given by energy density
which serves as label)
of the quasi-particle model. Such a family is exhibited in Fig.~\ref{fig:EoS}.
Surprising is the similarity of QPM(1.0) and the bag model EoS used in \cite{Kolb}.
Equipped with such a QCD based EoS one can compare results of hydrodynamical calculations
with data, e.g., for the azimuthal anisotropy $v_2$. 
Examples are exhibited in Fig.~\ref{fig:v2} for 
$\Lambda, \Xi, \Omega, \phi, D$.

\end{document}